\documentstyle[aps,preprint]{revtex}

\author{Diego F. Torres
\thanks{Electronic address: dtorres@venus.fisica.unlp.edu.ar}}
 
\address{Departamento de F\'{\i}sica, Universidad Nacional de La Plata,\\
C.C. 67, C. P. 1900, La Plata, Buenos Aires, Argentina}

\title{Boson Stars in General Scalar-Tensor Gravitation: Equilibrium
Configurations}

\begin{document}
                                                 
\maketitle
\begin{abstract}

We study equilibrium configurations of boson stars 
in the framework of general scalar-tensor theories of gravitation.
We analyse several possible couplings, with acceptable weak field 
limit and, when known, nucleosynthesis bounds, 
in order to work in the cosmologically 
more realistic cases of this kind of theories.
We found that for general scalar-tensor gravitation, the range of masses
boson stars might have is comparable with the general relativistic case.
We also analyse the possible formation of boson stars along different eras
of cosmic evolution, allowing for the effective gravitational constant
far out form the star to deviate from its current value. 
In these cases,
we found that the boson stars masses are sensitive to this kind of variations,
within a typical few percent.
We also study cases in which the coupling is 
implicitly defined, through the dependence on the
radial coordinate,  allowing  it to
have significant  variations in the radius of the structure.

{\it PACS number(s):} 04.50.+h, 04.40Dg, 95.35.+d

\end{abstract}

\newpage
\section{Introduction}

Boson stars, stellar structures first proposed by Ruffini and Bonazzola \cite{RB},
are gravitationally bound macroscopic quantum states made up of scalar
bosons. 
They differ from neutron stars, their fermionic counterparts, in that their
pressure support derives from the uncertainty relation rather than Pauli's exclusion
principle.
Although the seminal work of reference \cite{RB} was published in 1969,
it was followed up only in the past decade. In these recents years, cosmology
has been refurnished with the introduction of several ideas concerning the
critical role scalar field may would have in the evolution of the universe.
This revived the possibility of constructing stellar objects made up of these
scalars instead of conventional fermions.

Considering bosons as described by a non-interacting, massive, complex
scalar field, Ruffini and Bonazzola solved the equations of motion given by
the Einstein
field equations and the Klein-Gordon equation.
The general setting they used is the same as the one explained below. 
They found that the masses of such boson stars were of order $M \simeq
M_{pl}^2/m$, where $M_{pl}$ is the Planck mass and $m$ is the boson
mass. This model served to open the possibility that boson
stars might indeed exist in nature, although their masses
were small enough as to discard them as a viable solution
to the dark matter problem. The order of magnitude
of these boson stars coincides with simple computations.
For a quantum state confined into a region of radius $R$, and with
units given by $h=c=1$, the boson momemtum is $p=1/R$.
If the star is moderately relativistic $p\simeq m$, then $R \simeq 1/m$.
If we equate $R$ with the Scharszchild radius $2M_{pl}^{-2}M$
(recall that $G=M_{pl}^{-2}$) we get $M \simeq M_{pl}^2 /m$.
Later work made by Colpi, Shapiro and Wasserman \cite{CSW}
introduced a self-interaction term for the scalar field. With this addendum,
stellar equilibrium configurations had masses of order $M_{pl}^3/m^2$,
which are of the same order than the Chandrasekhar mass ($\simeq 1$ solar
masses). The stability of these objects have also been analysed with similar
results to those of the neutronic case \cite{stability}.
Taking Ruffini and Bonazzola and Colpi et. al. works as starting points, 
several extensions
have been proposed. Jetzer and van der Bij \cite{U(1)} considered
the inclusion of a U(1) gauge charge and Jetzer \cite{U(1)s} studied then
its stability properties. Non-minimally couplings for the scalars have also been
analysed in \cite{NM}. These and other related models are reviewed in \cite{PRBS}.
In addition, provided many of this objects may would have a primordial
origin, while being formed in a gas of boson and fermions, it is expected
that boson-fermions stars might exist. This was studied by Henriques et. al.
\cite{BF} without interaction and, recently, by de Sousa et. al. with 
current-current type interaction \cite{deSousa}. Finally, the possible
understanding of galactic halo properties by means of boson stars
models have also been proposed \cite{HALOS}.                             

Boson stars solutions have been, however, scarcely analysed in the framework 
of alternative theories of gravitation. We are particularly interested in 
scalar-tensor gravity, in which the effective gravitational constant is a 
field variable \cite{ST}.
Historically, most interest has been given to Brans-Dicke (BD) gravity,
in which the coupling function $\omega(\phi)$, 
free function these theories have,  is constant. To ensure
that the weak field limit of this theory agrees with current observations,
$\omega$ has to be big enough \cite{WILL}. But in general,          
when $\omega$  varies, we need that
$\omega \rightarrow \infty$ and $\omega^{-3} 
d\omega/d\phi \rightarrow 0$
when $t \rightarrow \infty$, to allow the weak field limit 
of scalar-tensor gravitation
accord well with general relativity (GR) tested predictions. Afterwards, 
soon was
realized that these general scalar-tensor theories would admit 
significant deviations from GR in the past \cite{ST}
and that they could be a useful tool in the understanding
of early universe models.  The interest on them was recently rekindled by
inflationary scenarios \cite{INF} and fundamental theories that seek 
to incorporate gravity with other forces of nature \cite{STRINGS}.
In general, almost all studies made on scalar-tensor gravitation
focus in the cosmological models they lead. This is in order to put several
constraints upon the coupling function.
Observational bounds, mainly coming from weak field tests \cite{WILL}
and nucleosynthesis\cite{NUCLEO1,NUCLEO2,NUCLEO3,TORRES_N} are more
restrictive if exact analytical solutions are known for the cosmological
equations. A few years ago, Barrow \cite{BARROW},
Barrow and Mimoso \cite{BARROW-MIMOSO} and Mimoso and Wands
\cite{MW} derived algebraic numerical methods that allow 
Friedmann-Robertson-Walker (FRW) solutions to be found in models with
matter content in the form of a barotropic fluid for any kind of coupling 
$\omega(\phi)$. Some of these methods were recently extended to
incorporate non-minimally coupled theories, even in the cases 
in which the functions
involved in the lagrangian do not posses analytical inverse 
\cite{TORRES_V,TORRES_S,TORRES_A}. 
This extensions showed the possibility of classify
the cosmological behaviour of scalar-tensor theories in equivalence sets,
where the field itself is a class variable.

In an astrophysical setting, if a scalar-tensor theory describes
gravitation, the value of the effective gravitational constant far out
from the star must not necessary be the Newton constant, but the value given by 
the evolution of a cosmological model at the time of formation of the stellar
object. As we shall see below, this may change the boundary conditions of 
the problem. 
Within this gravitational framework, boson stars where
analysed only in the simplest case. Gunderson and Jensen \cite{Gunderson}
addressed the possible existence of such objects in Brans-Dicke gravity,
with and without self-interaction. They adopted a fixed boundary condition
for the field equal to 1 --dimensionless Newton constant-- and
found that in general, for almost all
$\omega$, the Brans-Dicke model of 
boson stars gives a maximum mass smaller than the general relativistic
model in a few percent \cite{Gunderson}.
A similar work addressed the existence of boson stars in a gravitational
theory with dilaton \cite{DIL} and its results
coincides with the previous case. 


The aim of this work is then, to present a comprehensive study on the
possible existence of
boson stars solutions in {\it general scalar-tensor theories}. In the case
of their existence, we want to analyse the values they give for masses 
of typical objects
and others dynamical variables of interest, like the typical behavior
of the scalar.
We also
want to see if  modifications
in the boundary condition for the Brans-Dicke scalar --due
to cosmological evolution-- produce any noticeable deviation
in the masses of equilibrium configurations. 
We also implicitly define the coupling in order to allow 
sustantive variations of it in the radius of the structure.
We then study equilibrium configurations in these schemes.

The rest of the paper is organized as follows. In Sec. II we introduce 
the formalism
for boson stars construction together with the numerical recipes used.
In Sec. III we make choices of coupling functions and in Sec. IV the results
for them are presented. The last section deals with our conclusions.

\section{Formalism}

\subsection{Gravitational theory and boson system}

We first derive the equations that corresponds to the general 
scalar-tensor theory. The action for this kind of generalized BD
theories is

\begin{equation}
\label{action}
S=\frac{1}{16\pi }\int \sqrt{-g}dx^4\left[ \phi R-\frac{\omega(\phi )}%
\phi g^{\mu \nu }\phi _{,\mu }\phi _{,\nu }+ 16\pi \L_m\right],
\end{equation}
where $g=Det \;g_{\mu\nu}$, $R$ is the scalar curvature, $\omega$ is the
coupling function and $L_m$ represents the matter 
content of the system. We take this $L_m$ to be the lagrangian density of 
a complex, massive, self-interacting scalar field. 
This lagragian reads as:

\begin{equation}
L_m=-\frac{1}{2} g^{\mu\nu} \psi^*_{,\mu} \psi_{,\nu} -\frac{1}{2} m^2 |\psi|^2 
-\frac{1}{4} \lambda |\psi|^4 .
\end{equation}

Varying the action with respect to the dynamical variables $g^{\mu\nu}$ and
$\phi$ we obtain the field equations:

\begin{equation}
\label{field0}
R_{\mu\nu}-\frac{1}{2} g^{\mu\nu}R=\frac{8\pi}{\phi}T_{\mu\nu}
+\frac{\omega(\phi )}{\phi} \left( \phi_{,\mu} \phi_{,\nu}-
\frac{1}{2} g_{\mu \nu }\phi^{,\alpha}\phi_{,\alpha}\right)+ 
\frac{1}{\phi} \left( \phi_{,\mu;\nu} -g_{\mu\nu} \Box\phi \right)
\end{equation}

\begin{equation}
\label{field00}
\Box \phi =\frac{1}{2\omega+3} \left[ 8\pi T-\frac{d\omega}{d\phi}
\phi^{,\alpha}\phi_{,\alpha}\right],
\end{equation}
where we have introduced $T_{\mu\nu}$ as the energy-momentum tensor for
matter fields and $T$ as its trace. This energy-momentum tensor is given by

\begin{equation}
\label{emt}
T_{\mu\nu} =\frac{1}{2} \left( \psi^*_{,\mu }\psi_{,\nu } +\psi_{,\mu }
\psi^*_{,\nu }\right) -\frac{1}{2} g_{\mu\nu} \left( g^{\alpha\beta}
\psi^*_{,\alpha }\psi_{,\beta} +m^2 |\psi|^2 + \frac{1}{2} \lambda |\psi|^4 
\right).
\end{equation}
The covariant derivative of this tensor is null. That may be proved either from 
the field equations, recalling the Bianchi identities, or by intuitive arguments
since the minimally coupling between the field $\phi$ and the matter 
fields. This implies:

\begin{equation}
\label{fieldb}
\psi^{,\mu}_{,\mu}-m^2 \psi -\lambda |\psi|^2  \psi^*=0.
\end{equation}

We now introduce the background metric. That is the corresponding to
a spherically symmetric system, because of the symmetry we impose
upon the star. Then:

\begin{equation}
\label{metric}
ds^2=-B(r) dt^2 + A(r) dr^2 +r^2 d\Omega^2.
\end{equation}
We also demand a spherically symmetric form for the field which describe the
boson, {\it i.e.}, we adopt:

\begin{equation}
\label{boson}
\psi(r,t)=\chi(r) \exp{[-i\varpi t]}.
\end{equation}
Semiclassically, we are able to think about $\chi$ expanded in creation
and annihiliation operators and $T_{\mu\nu}$ as 
an expectation value in a given 
configuration with a large number of bosons \cite{RB}.
Using the metric (\ref{metric}) and the equation defining the form
of the boson field (\ref{boson}) together with the energy-momemtum
tensor (\ref{emt}) in the field equations (\ref{field0},\ref{field00}) we get the 
equations of structure of the star. Before we explicitly write  them
we are going to introduce a rescaled radial coordinate by:

\begin{equation}
\label{x}
x=mr.
\end{equation}
From now on, a prime will denote a differentiation with respect to the variable
$x$. We also define dimesionless quantities by

\begin{equation}
\label{dimensionless}
\Omega=\frac{\varpi}{m}, \;\;\;\; \;  \Phi=\frac{\phi}{M_{pl}^2} ,\;\;\;\;\;
\sigma=\sqrt{4\pi} \chi(r),\;\;\;\;\;and\;\;\;\;\;
\Lambda=\frac{\lambda}{4\pi} \left( \frac{M_{pl}}{m} \right)^2,
\end{equation}
where $M_{pl}$ is the Planck mass. In order to consider the total
amount of mass of the star within a radius $x$ we change the function
$A$ in the metric to its Schwarzschild form:

\begin{equation}
\label{M}
A(x)=\left(1-\frac{2M(x)}{x}\right)^{-1}.
\end{equation}
Then, the total mass will be given by $M(\infty)$ and will corresponds to

\begin{equation}
M_{star}= \frac {M(\infty)}{m} M_{pl}^2,
\end{equation}
for a given value of $m$.
With all these definitions, the equations of structure reduce to the following 
set:

\begin{equation}                     
\label{field1}
\sigma^{\prime \prime} + \sigma^{\prime} \left( \frac{B^\prime}{2B} -
\frac{A^\prime}{2A} + \frac{2}{x} \right) + A \left[ \left(
\frac{\Omega^2}{B}-1 \right)\sigma - \Lambda \sigma^3 \right]=0
\end{equation}

\begin{equation}
\label{field2}
\Phi^{\prime \prime} + \Phi^{\prime} \left( \frac{B^\prime}{2B} -
\frac{A^\prime}{2A} + \frac{2}{x} \right) - \frac{2A}{2\omega+3} 
\left[ \left(
\frac{\Omega^2}{B}-2 \right)\sigma^2 -\frac{\sigma^{\prime 2}}{A} - 
\Lambda \sigma^4 \right] + \frac{1}{2\omega+3}\frac{d\omega}{d\Phi} 
\Phi^{\prime 2}=0
\end{equation}

\begin{eqnarray}
\label{field3}
\frac{B^{\prime}}{xB}-\frac{A}{x^2}\left( 1-\frac{1}{A} \right)=
\frac{A}{\Phi} 
\left[ \left(
\frac{\Omega^2}{B}-1 \right)\sigma^2 +\frac{\sigma^{\prime 2}}{A} - 
\frac{\Lambda}{2} \sigma^4 \right] 
+ \frac{\omega}{2}\left(\frac{\Phi^{\prime}}{\Phi}\right)^2 -
\frac{A}{\Phi} \frac{2}{2\omega+3}\times \nonumber \\ 
\left[ \left(\frac{\Omega^2}{B}-2 \right)\sigma^2 -\frac{\sigma^{\prime 2}}{A} - 
\Lambda \sigma^4 \right] +   
\left( \frac{\Phi^{\prime \prime}}{\Phi} - \frac{1}{2}\frac{\Phi^{\prime}}{\Phi} 
\frac{A^\prime}{A} \right) + 
\frac{1}{2\omega+3}\frac{d\omega}{d\Phi} 
\frac{\Phi^{\prime 2}}{\Phi}
\end{eqnarray}

\begin{eqnarray}
\label{field4}
\frac{2BM^\prime}{x^2}=\frac{B}{\Phi} 
\left[ \left(
\frac{\Omega^2}{B}+1 \right)\sigma^2 +\frac{\sigma^{\prime 2}}{A} + 
\frac{\Lambda}{2} \sigma^4 \right] 
+ \frac{\omega}{2}\frac{B}{A}\left(\frac{\Phi^{\prime}}{\Phi}\right)^2+ 
\frac{B}{\Phi} \frac{2}{2\omega+3} \times \nonumber \\
\left[ \left(\frac{\Omega^2}{B}-2 \right)\sigma^2 -\frac{\sigma^{\prime 2}}{A} - 
\Lambda \sigma^4 \right] -
\frac{B}{A(2\omega+3)}\frac{d\omega}{d\Phi} 
\frac{\Phi^{\prime 2}}{\Phi} - 
\frac{1}{2}\frac{\Phi^{\prime}}{\Phi} 
\frac{B^\prime}{A}.
\end{eqnarray}
It is important to note that these equations reduce to the known
BD ones of reference \cite{Gunderson} when $\omega$ is taken as a constant
and to those of GR of reference \cite{CSW} when $\Phi \rightarrow \Phi_0$,
constant.

\subsection{Numerical procedure and boundary conditions}

We shall carry out a numerical integration from the centre of the star
outwards towards radial infinity. The boundary condition for the system 
are the following. Concerning $\sigma$, we require a finite mass, 
which implies $\sigma(\infty)=0$ and non-singularity 
at the origin, {\it i.e.} $\sigma(0)$ a finite 
constant and $\sigma^\prime (0)=0$. We shall look for zero node
solutions because, as remarked in \cite{CSW}, it is reasonable to
suppose them as the lowest energy bound states. We shall demand
asymptotic flatness, which means $B(\infty)=1$ and $A(\infty)=1$,
this last condition, ensured by the equations themselves.
Non-singularity at the origin also requires $M(0)=0$. Finally, $\Phi(\infty)$
must take the value of $\Phi$ in an appropiate cosmological model
at the time of stellar formation. If not otherwise specified, 
it will be considered
as $1$. Note that this boundary condition differs from the others
in being less restrictive, in fact, preliminarily,
it impose only an {\it acceptable boundary value prescription}.
We shall manage this boundary condition by 
stopping the integration at the asymptotic region, where 
$\Phi=\Phi(\infty),\; (\pm 10^{-6})$, and the derivative $\Phi^\prime (\infty)$ 
tends to zero.
All these boundary conditions generate an eigenvalue problem
for $\Omega$. In order to abtain accurate results,
this eigenvalue has to be specified with at least seven significant
figures in a typical case, this is within the capability of a double precision
numerical method and was also the case in general relativistic cases \cite{BF}.
Note that due to the form of the equations, which are
linear in $B$, we can integrate the system without impose the boundary
condition on $B$ from the beginning. Instead, we ultimately rescale
$B$ and $\Omega$ in order to satisfy flatness requirements.

The numerical method  we shall use is a fourth order Runge-Kutta
and is based in the recipes of \cite{RECIPES}. Some of the subroutines
were modified in order to test the possibility of satisfying 
the boundary conditions
at each step of the integration (see \cite{RB} for details on this) and some
others were built in order to search for the eigenvalue and satisfy the
specific boundary conditions.
The program was tested in
the limiting cases of equations (\ref{field1}-\ref{field4}), {\it i.e.} GR and BD,
and agreement was found with reported results.

\section{The coupling function}              

As there is no {\it a priori} prescription about the form or the value
of $\omega$, we are interested in ascertaining the general behaviour
displayed by a wide range of scalar-tensor stars. The first Group of
couplings we are going to analyse have the property of tending to
infinity as $\phi \rightarrow \phi_0$, where $\phi_0$ may be taken as
the present value $\phi(t)$ or, equivalently, as the inverse of the Newton
constant.  We shall take the forms that Barrow and Parsons recently
analysed in  a cosmological setting \cite{BP}. They are:

\begin{itemize}
 
\item Theory 1.  $2\omega+3=2B_1 |1-\phi/\phi_0|^{-\alpha}$, with
$\alpha >0$ and $B_1>0$ constants.
                               
\item Theory 2.  $2\omega+3=2B_2 \ln|\phi/\phi_0|^{-2\delta}$, with
$\delta >0$ and $B_2>0$ constants.
                               
 \item Theory 3.  $2\omega+3=2B_3 |1-(\phi/\phi_0)^{\beta}|^{-1}$, with
$\beta >0$ and $B_3>0$ constants.

\end{itemize}
The behavior of these theories in a FRW metric
was analitically studied in \cite{BP} and weak field limit
constraints upon the parameters were provided there also.
Note that, while using these couplings in our equations,
the functions dependences shift from cosmic time to radial
coordinate. Cosmological solutions for this group allows $\phi$ approach to
$\phi_0$ from below, {\it i.e.} $\phi \in (0, \infty)$ or from above,
$\phi \in (\infty,0)$. This implies that the boundary condition
in $\Phi$ may be equal to, less than or bigger than $1$.   
Theories 1.--3. approach BD when $\Phi \rightarrow 0$ and to next theory
(Theory 4.) when $\Phi \rightarrow \infty$. Note also that  the 
weak field constraints
are, in fact, independent of the form of the cosmological solutions
provided $\Phi \rightarrow \Phi_0$, when $t$ is big enough. It is this latter
requirement which 
introduce further restrictions upon the the parameter space,
specially in the exponents, which vary as a function of the cosmic era  
\cite{BP}.

The second Group of coupling functions will be represented by:

\begin{itemize}

\item Theory 4.  $2\omega+3=\omega_0 \;\;\phi^n$, with
$n >0$ and $\omega_0$ constants.
\end{itemize}
It also has analytical solutions \cite{BARROW-MIMOSO} and even we know
nucleosynthesis bounds for it \cite{TORRES_N}. This group differs from the 
first in that, although growing with time, they only reach GR when $\phi 
\rightarrow \infty$, that is, when $t \rightarrow \infty$, $\phi$
do not tends to $\phi_0$. To
normalize we may set $\phi(t=today)=1$.

Finally, the third Group we shall analyse consists of {\it local
implicitly defined functions} of the form:                    

\begin{itemize}

\item Theory 5.  $\omega=\omega(x)=\omega(x(\phi))$

\end{itemize}
The aim in doing so is explicitly see how can one manage the behavior of
$\omega$ within the radius of the star.
Note that these are implicit definitions, being ultimately necessary
to invert $\phi(x)$ to get the correct dependence of the coupling function.
If  $\phi(x)$ is a monotonous function, then, the existence of this inverse
is analitically ensured. It is worth recalling that the limit $x\rightarrow
\infty$ if of crucial importance. Far out from the star we would want to
recover a scalar-tensor theory with a cosmological well-behaved
evolution.

\section{Results}

\subsection{Group I and II Couplings}

Although formally and conceptually different, when considering 
a cosmological setting, the theories described as Group I and 
Group II resulted to be similar concerning boson stars solutions.
In addition, most of the similarities arise inside Group I couplings,
where
simulations based on these theories suggest that anyone 
can be mapped into the 
other for convenient choices of each particular set of their free parameters.
Concerning Theory 4. some differences have to be remarked, and so we do
below, but its general behaviour do not differ very much from Group I
couplings.
Taking this into account, we shall present in deep detail
only the case of Theory 1. and we shall make some comments on 
others special situations.

We shall first consider boson stars based on Theory 1. gravitation.
We take $B_1=5$ and $\alpha=2$ and look for models in
equilibrium for different values of the central density and strengh
of the self-interaction. What we found is sketched in Fig.  1. Recall that 
the value
$\alpha=2$ is one extremum of the interval which admits
convergency to GR in  a cosmological evolution with
a perfect fluid model as matter source ($\gamma \in (0,4/3); 
p=(\gamma -1)\rho)$ \cite{BP}.
We can note from there that the general form of the graph is preserved when 
compared with both, GR and BD cases. The boson stars 
masses increase from the
BD case with $\omega=6$ presented by Gunderson and Jensen. 
In fact, results
for values of $\alpha$ greater than $1$ are extremely similar to those of
general relativity and the BD scalar is almost 
numerically constant along the boson
star structure. This is something that could be expected in the case solutions
may exist, because of the rapidly approaching scheme to GR that Theory 1.
develops when $\alpha$ is big enough.
Things change when considering values of $\alpha$ smaller than $1$.
Table \ref{Table1} presents computations for models with $B_1=5$ and $\alpha
=0.5$. Recall that this value of $\alpha$ is the smaller value that preserves
the weak field limit in a cosmological evolution and one of the extremum
which guarantees convergency to GR in the case of the radiation era \cite{BP}.
Note that the equilibrium configuration in each case always choice
a bigger value of $\Phi$ at the center of the star, which implies less
gravitationally 
bounded objects than those of GR. This was also the case of BD
models. The values of the masses are smaller than GR ones but still greater 
than the ones which do not behave as required for a cosmological
setting, as for instance the BD case with $\omega=6$.
So, Theory 1. has viable solutions for boson stars structures
where masses are compatible with simplest cases.
Concerning the behaviour of $\sigma$ as a function of $x$, it has the same
convexity properties commented for GR and BD models. 
Fig.  2. shows its
behaviour for typical values of the parameter space. The same does
Fig.  3. for the behaviour of $\Phi$. 
The dependences of the masses of the equilibrium structures upon the
parameters of the gravitational theory was tested in further detail. It was
found that for values of $\alpha$ grater than $1$, changes in $B_1$
do not produce noticeable changes in the mass. The oppossite happens for
smaller values of $\alpha$. Table \ref{Table2} represents these trends
in a more quantitative form for  $\Lambda=100$ and $\sigma(0)=
0.100$.    
Finally, we address the possible variation of $M(\infty)$ with a deviation
in the boundary condition for $\Phi$, the effective gravitational
constant far out from the star. This is aimed in getting a first insight
of possible boson stars formation along differents eras of cosmic evolution.
As Theory 1. admits cosmological solutions with values of $\Phi$ greater 
or smaller than $1$ we consider both cases as possible boundary conditions.
Table \ref{Table3} shows $M(\infty)$ and the corresponding $\Phi(0)$
value for each choice
of the boundary condition in three particulars models. It is interesting
to note that, in the first place, {\it  masses are sensitive to variations
in the boundary condition of the scalar within a few percent as typical
order of magnitude}, and second, {\it the behavior of models varies
with $\Lambda$}. If $\Lambda$ is big enough
(greater than 10) a growing mass appears with 
a growing boundary condition. Otherwise, the models show a peak
in the masses within the range explored for $\Phi(\infty)$.
                                     

Concerning Theory 4. it has to be noted that the parameter space
is not mainly constrained by weak field test \cite{BARROW-MIMOSO}, 
which do not limit the values of $n$ --provided $n\,>\,0$--, but it is by
nucleosynthesis processes \cite{TORRES_N}.
This bounds, which resulted in lower limits for $n$,
are provided once the cosmological parameters 
$\Omega_0$ and the Hubble constant $H_0$ are given.
A common characteristic of these Group is that masses of equilibrium
configurations are smaller than the cases previously studied and much more
smaller when compared with GR. A typical example is the $\Lambda=100$
and $\sigma(0)=0.100$ model. For $\omega_0=2$ and $n=3$, 
$M(\infty)=1.870$. Note, however, that this value of  $n$ 
produce acceptable nucleosynthesis consequences only in a range of
$\Omega_0 h^2 < 0.25$.
Another thing to note is that Theory 4. resulted in the ones
which more dependence on the paremeters show for small values
of $n$ and $\omega_0$. There, variations may reach a typical 
$10\%$ in mass.

\subsection{Group III Couplings}

The third and last group of couplings we shall analyse consists in implicitly
defined functions of the form of Theory 5. 
As an example we choose several forms of the couplings, results for them
can be seen in Table \ref{Table4}, for the model given by
$\Lambda=0$ and  $\sigma(0)=0.325$.
These functions are enough to get a feeling of which the idea is.
Locally there is no prescription upon $\omega$, while far out 
from the star we would want to recover a scalar-tensor theory 
cosmologically well-behaved. 
Scalar-tensor theories
which deviates more from GR are those which can be compared
with BD theories of small $\omega$. The choice of the different functions
is focused to encompass GR at the asymptotic region while admitting
severe deviation inside the structure.
For all cases, when $x\rightarrow \infty$, $\omega
\rightarrow \infty$,  making these theories
cosmologically acceptable. 
If not otherwise specified,  all cases present monotonous
$\Phi$-functions.
The correct dependence of the coupling, $\omega(\Phi)$, may
be lately obtained from the inverse of the function $x(\Phi)$.
It is worth recalling that
the dependence of $\omega$ with $\Phi$ changes whenever the
model change. For instance, in passing through different values 
of $\Lambda$, the functional form of $\Phi(x)$ change, and the
same does its inverse. This implies that even without changing 
$\omega=\omega(x)$ we are changing $\omega=\omega(\Phi)$.
For the cases studied in Table \ref{Table4} we found that the order of
magnitude of boson stars masses remains the same, although 
some cases with very small masses may arise --also in the cases
of $\Lambda \neq 0$. These, in general,  mild variations in the boson
star properties must be explained in terms of the complex structure
of the differential system.
The terms proportional to the derivatives of the couplings
are also proportional to the derivative of $\Phi$, which in turn
must be obtained from the solution of the system.

\section{Conclusions}
                 
In the last few years, the possibility of  constructing
complete cosmologies, by encompassing exact analytical
solutions of general scalar-tensor gravitation, have raised
an enourmous interest on these kind of theories, which has to be
added to the developed by the applicability of them to inflationary
scenarios. Once the cosmological setting is fixed, we have to analyse
possible astrophysical consequences of having, for instance, a different
value for the gravitational constant or a different rate of expansion.
As an example, we should mention
a recent work about primordial formation and evaporation of
black holes \cite{BARROW-CARR}.

In this work, we have analysed the possible existence of boson
stars solutions within the framework of general scalar-tensor 
theories. It then extends a previous paper by Gunderson and Jensen
\cite{Gunderson} were solutions to the Brans-Dicke equations
were considered. 
We have shown the theoretical 
construction of such systems in general cases of alternative
gravity, all them contrasted with the gravity tests known
up to date.  Different kind of couplings with exact cosmological solutions
and others that allow a significant  variation within the radius of the 
star were considered. 
We found that the order of magnitude of the general relativistic 
boson star masses
do not vary when these more realistic cases of scalar-tensor gravity are 
the basis of the gravitational theory.
In general, and because general forms of couplings can be expanded
in the form of a series of Group I and/or Group II couplings, we may state
that boson stars might exist for any of these gravitational settings. 
We also found an interesting situation concerning the evolution
of boson stars masses as a function of the time of formation
of the stellar object. It appreciable vary, within a typical
few percent, when cosmological time scales are considered.
If this fact may provide a useful basis for searching new
observational consequences and/or bounds upon the coupling
function is currently under study.    
Finally it remains to be considered the question of stability,
for which we do not know results, even in the Brans-Dicke case.
We hope to report on it in a forecoming work.

\subsection*{Acknowledgments}

It is the author's pleausure to acknowledge I. Andruchow, S. Grigera,
T. Grigera and, specially, O. Benvenuto for their help with
the numerical procedure and H. Vucetich for useful comments 
and a critical reading of the manuscript. 
The author also aknowledge D. Krmpotik and R. Borzi
for the use of computing facilities of 
{\it Office 23} at UNLP and partial support by 
CONICET.

\begin{table}
\caption{Boson stars masses for Theory 1. with $B_1=5$ and $\alpha=0.5$.}
\begin{tabular}{|ccccc|}
$\Lambda$      &        $\sigma(0)$     &      $B(0)$  & $\Phi(0)$      &      
$M(\infty)$ \\   \hline
0 &  0.325  &  0.4231 & 1.0007 & 0.627  \\
10 & 0.225 & 0.4163 & 1.0010  &  0.919 \\
100   &   0.100  &  0.3853  & 1.0011 & 2.248\\ 
200    &  0.070 &  0.4256  & 1.0009  & 3.128\\   
\end{tabular}
\label{Table1}
\end{table}

\begin{table}
\caption{Dependences of boson stars masses upon the parameter
space for Theory 1. with
$\Lambda=100$, $\sigma(0)=0.100$.}                                  
\begin{tabular}{|cccc|}
 $B_1$  & $\alpha$   & $\Phi(0)$ &     $M(\infty)$ \\   \hline
2 &  0.5   &     1.0053   & 2.245  \\
5 & 0.5 & 1.0010      &  2.248  \\
 8   &   0.5  &  1.0004 & 2.249\\ 
2    &  1.0 &  1.0000 & 2.250 \\  
8 & 1.5 & 1.0000&   2.250 \\
8 & 2.0 & 1.0000 &  2.250 \\        
\end{tabular}
\label{Table2}
\end{table}

\begin{table}
\caption{Boson stars masses as a function of the boundary condition for $\Phi$.
First set shows the model with $\Lambda=100$ and $\sigma(0)=0.100$,
the second set shows: $\Lambda=10$ and $\sigma(0)=0.225$, while the third,
$\Lambda=0$ and $\sigma(0)=0.325$. These models 
are for Theory 1. with $B_1=5$ and $\alpha=0.5$.}
\begin{tabular}{|c|cc|cc|cc|}
$\Phi(\infty)$      &        $\Phi(0)$     &      $M(\infty)$  & $\Phi(0)$      &      
$M(\infty)$ &    $\Phi(0)$     &      $M(\infty)$\\   \hline
0.90 &  0.9127  &  2.096 &    0.9128 & 0.875 &    0.9111 & 0.610  \\
0.95  &  0.9593 & 2.164 &  0.9593 & 0.893 &               0.9581 & 0.616 \\
 1.00   &   1.0009  &  2.253  &     1.0010 &  0.920 & 1.0007 & 0.627\\ 
1.05    &  1.0615 &  2.263  & 1.0617 & 0.916 & 1.0600  & 0.618\\ 
1.10 & 1.1167 & 2.295 & 1.1170 & 0.921 & 1.1145 & 0.614 \\
\end{tabular}
\label{Table3}
\end{table}

\begin{table}
\caption{Masses of boson stars for implicitly defined scalar-tensor theories. 
Results for the model
$\Lambda=0$ and $\sigma(0)=0.325$ are shown. 
A small star point that the $\Phi$-function is not monotonous, typically
in the innermost region. Boundary condition on the BD scalar was setted
equal to 1 although deviations provided by an asymptotic
derivative of $\Phi$
of order $10^{-4}$ were accepted.}
\begin{tabular}{|ccc|}
Theory 5.: $\omega=\omega(x)$  &  $M(\infty)$      &      $\Phi(0)$ \\   \hline
0.1 x &  0.538   & 1.1011  \\
10 x & 0.624 & 1.0075  \\
$\log(x)^*$ & 0.577 & 1.0754 \\
$\exp(0.01 x)$ & 0.539 & 1.0760 \\
\end{tabular}
\label{Table4}
\end{table}

\begin{figure}  
\label{fig.1}
\caption{Boson stars masses of Theory 1. for $B_1=5$ and $\alpha=2$
and different values of  $\Lambda$ and  $\sigma(0)$. There are 34 models
for each value of $\Lambda$. Numerical values in this graph are very similar 
to the ones 
derived for General Relativity boson stars masses.}
\end{figure}

\begin{figure}  
\label{fig.2}
\caption{Behaviour of $\sigma$ as a function of the radial coordinate for two
typical models of scalar-tensor boson stars; Theory 1. 
with $B_1=5$ and $\alpha=0.5$.}
\end{figure}

\begin{figure}  
\label{fig.3}
\caption{Behaviour of $\Phi$ as a function of the radial coordinate for two
typical models of scalar-tensor boson stars; Theory 1.
with $B_1=5$ and $\alpha=0.5$.}
\end{figure}

\end{document}